\documentclass[%
reprint,
showpacs,showkeys,
amsmath,amssymb,
aip,
jvsta,
]{revtex4-1}

\usepackage{graphicx,subfigure,float}% Include figure files
\usepackage{adjustbox}
\usepackage{dcolumn}% Align table columns on decimal point
\usepackage{bm}% bold math
\usepackage{color}
\usepackage{hyperref}% add hypertext capabilities
\hypersetup{colorlinks=true,citecolor=blue%,urlcolor=cyan
}
\begin{document}

%\preprint{APS/123-QED}

\title{Effect of substrate bias on microstructure of epitaxial film grown by HiPIMS: An atomistic simulation}
\author{Movaffaq Kateb}
\email[Corresponding author email address: ]{movaffaqk@ru.is}
\affiliation{Science Institute, University of Iceland,
Dunhaga 3, IS-107 Reykjavik, Iceland}
\affiliation{School of Science \& Engineering, Reykjavik University, Menntavegur 1, IS-102 Reykjavik, Iceland}

\author{Jon Tomas Gudmundsson}
\affiliation{Science Institute, University of Iceland,
Dunhaga 3, IS-107 Reykjavik, Iceland}
\affiliation{Department of Space and Plasma Physics, School of Electrical Engineering and Computer Science, KTH Royal Institute of Technology, SE-100 44, Stockholm, Sweden}

\author{Snorri Ingvarsson}
\affiliation{Science Institute, University of Iceland,
Dunhaga 3, IS-107 Reykjavik, Iceland}

\date{\today}

\begin{abstract}
We explore combination of high power impulse magnetron sputtering (HiPIMS) and substrate bias for the epitaxial growth of Cu film on Cu (111) substrate by molecular dynamics simulation. A fully ionized deposition flux was used to represent the high ionization fraction in the HiPIMS process. To mimic different substrate bias, we assumed the deposition flux with a flat energy distribution in the low, moderate and high energy ranges. We also compared the results of fully ionized flux with results assuming a completely neutral flux, in analogy with thermal evaporation. It is confirmed that in the low energy regime, HiPIMS presents a slightly smoother surface and more interface mixing compared to that of thermal evaporation. In the moderate energy HiPIMS, however, an atomically smooth surface was obtained with a slight increase in the interface mixing compared to low energy HiPIMS. In the high energy regime, HiPIMS presents severe interface mixing with a smooth surface but limited growth due to resputtering from the surface. The results also indicate that fewer crystal defects appear in the film for moderate energy HiPIMS. We attribute this behavior to the repetition frequency of collision events. In particular high energy HiPIMS suffers from high repetition of collision events which does not allow reconstruction of the film. While in the low energy HiPIMS there are not enough events to overcome island growth. At moderate energy, collision events repeat in a manner that provides enough time for reconstruction which results in a smooth surface, fewer defects and limited intermixing.

%\begin{description}
%5Secondary publications and information retrieval purposes.
%\end{description}
\end{abstract}

% sputtering, molecular dynamic, plasma ionization, plasma neutrals, plasma simulation

\maketitle

%\tableofcontents

\section{Introduction}
Among the variety of ionized physical vapor deposition (PVD) techniques, high power impulse magnetron sputtering (HiPIMS) has attracted significant attention over the past few decades. \citep{helmersson06:1,gudmundsson12:030801} In particular, HiPIMS enables achieving discharge of high electron density through pulsing the cathode target to a high power density with unipolar voltage pulses, at low duty cycle, and low repetition frequency. \citep{helmersson06:1,gudmundsson12:030801,gudmundsson10:1360} In dc magnetron sputtering (dcMS), high power causes a considerable thermal load on the target which limits the plasma density to 10$^{15}-10^{17}$~m$^{-3}$ range \citep{seo04:1310,seo06:256,sigurjonsson08:062018} and leads to a low ionization density fraction in the deposition flux ($<$10~\%). \citep{christou00:2897} Thus, the majority of the film forming particles arriving at the substrate surface are electrically neutral and cannot be affected by a substrate bias. In HiPIMS, the target overheating problem is solved by applying high power impulses with a low duty cycle. As a result a peak electron density of 10$^{19}$~m$^{-3}$ can be achieved in the vicinity of the cathode target. \citep{gudmundsson02:249,bohlmark05:346,meier18:035006} This shortens the ionization mean free path and increases the ionization probability of sputtered atoms by collisions with energetic electrons. For instance, the ionization fraction of the sputtered flux from a Cu target has been reported to reach up to 70\% using HiPIMS. \citep{kouznetsov99:290} As a result HiPIMS presents not only smoother, \citep{magnus11:1621,sarakinos07:2108} denser \citep{samuelsson10:591} and more void-free \citep{alami05:278} coatings compared to dcMS, but it also allows for efficient control over film properties through applying a bias voltage to the substrate. \citep{hajihoseini2018} By applying a substrate bias the bombarding energy of the ions of the film forming material can be controlled. This provides an effective method to control properties such as the film texture and grain size. This can be performed by not only dc bias but also by synchronizing the substrate bias to the power supply. \citep{zednek20:49,greczynski19:060801} It was demonstrated earlier that the appearance of working gas and metal-ions at the substrate are separated in time. \citep{macak00:1533} This allows for metal-ion-synchronized substrate biasing.

In spite of extensive experimental efforts to study HiPIMS deposition with biased substrate, \citep{nedfors18:031510,hajihoseini2018,jablonka18:043103} the atomistic mechanisms that contribute to this process have not been studied so far. For instance, using X-ray diffraction, variation in the lattice parameter with the substrate bias voltage for VN films prepared by HiPIMS is observed and found to be associated with the nitrogen content of the films. \citep{hajihoseini2018} However, the mechanism that reduces nitrogen content and consequently changes lattice parameter, still remains unknown. Also a change in preferred crystal orientation (texture) is observed as the bias voltage is increased. \citep{nedfors18:031510,hajihoseini2018} Understanding such phenomena requires atomistic resolution and time scales that are covered by Monte Carlo (MC) \citep{muller1985,muller1986,muller1986jap,dodson1990} and molecular dynamics (MD) simulations. \citep{muller1987prb,muller1987} 

MD simulation has been widely utilized to study evaporation deposition in absence of ions. The results of these simulations can be classified within three categories, based on the energy of the incident atoms: (i) low energy (0.1 -- 2~eV) regime as a mimic of molecular beam epitaxy \citep{zhou1997}, (ii) moderate energy (2 -- 10~eV) for evaporation and (iii) high energy (10 -- 40~eV) supposedly representing dcMS deposition (without introducing ionic species). \citep{sprague1996} The films deposited at low energy conditions presented a porous and columnar microstructure \citep{muller1985,henderson1974,kim1977} which could become drastic in oblique deposition. \citep{henderson1974,kim1977,zhou1997,hubartt2013} It has also been shown that increased substrate temperature \citep{schneider1985,schneider1987prb,smith1996,zhou1997,zhang1998} and/or increased adatom energy \citep{smith1996,zhou1997,zhang1998} leads to a more defect-free film. In the moderate energy deposition, a competition between island growth and layer-by-layer growth was observed in which the latter became dominant as the energy of incident atoms approaches 10~eV. \citep{gilmore1991,zhang1998} The high energy case, however, presents mixing at the film-substrate interface. \citep{sprague1996,zhang1998,lugscheider1999}

On the other hand there have been efforts to model ion-assisted PVD i.e.\ a deposition flux consisting of both neutral adatoms and ions of the noble (working) gas. \citep{muller1986,muller1986jap,muller1987prb,muller1987} It is found that moderate energy adatoms and high energy ion bombardment cause densification of the film and a smooth surface. \citep{muller1987prb,muller1987,fang1993} Also higher ratio of ions to neutrals decreases the number of voids and reduces the surface roughness. \citep{muller1986} It has also been shown that ion bombardment can cause grain growth in the favorite crystal orientation (texture refinement). \citep{dong1998,dong1999} In addition, ion-assisted deposition enables a more uniform deposition on substrates with complex geometries. \citep{hwang2002,hwang2003} Furthermore, it has been shown that, the ion energy has a major effect on the surface roughness compared to ion incident angle. \citep{su2006} 

Most of the previous studies assumed a monodispersed energy for the flux, rather than an energy distribution. This might be a reasonable assumption in thermal evaporation but in order to model realistic ionized PVDs an energy distribution is necessary. \citep{bohlmark06:1522} For instance, it has been shown that a mixed flux of low and high energy adatoms, generates localized amorphization which was not observed in the above mentioned studies. \citep{houska2014} It has also been shown that a distribution function is necessary in order to model the deposition of amorphous nanocatalysts. \citep{xie2014} We have recently demonstrated simulations of deposition assuming a flat energy distribution with 0 -- 100\% ionized flux (for adatoms, not ions of the working gas). \citep{kateb2019} In partially and fully ionized systems, a localized amorphization followed by recrystallization has been observed. This occurs through the so called \emph{bi-collision event} which is responsible for reducing surface roughness and densification of the film. We would like to remark that, previous studies based on the energy distribution could only model the amorphization but not the recrystallization. \citep{houska2014,xie2014} For the first time, we observed generation of interstitial defects during the deposition that represents compressive stress in the film. Note that earlier studies of ion-assisted PVD were only able to demonstrate zero or tensile stress due to presence of vacancies and voids. \citep{muller1987,fang1993} Also we did not detect any interstitials in the film deposited using a neutral flux along with flat energy distribution. \citep{kateb2019} These examples indicate that defining an energy distribution alone is not sufficient for modeling a realistic ionized PVD and that ionization fraction in the flux must also be taken into account.

It is worth mentioning that in general the time scale that is achieved in MD simulation is on the order of tens of ns. Thus it cannot capture the entire HiPIMS pulse which is on the order of tens and hundreds of $\mu$s. Besides, the deposition rate in MD simulation is several orders of magnitude higher than for typical experimental conditions. However, during the pulse on time, the deposition rate of HiPIMS is one or two order of magnitude higher than for dcMS (cf. \citep{kateb2018}). Furthermore, the flux of ions is not constant during the pulse and is different for metal ions and working gas ions. \citep{greczynski19:060801} Combining these facts one can assume several orders of magnitude higher deposition rate to occur at the peak intensity of ion flux.

The aim of the present study is to understand the effect of substrate bias on the HiPIMS deposition. We assume roughly low, moderate and high energy distribution of the deposition flux to mimic different substrate bias  voltages. This is also of fundamental importance in order to understand synchronized bias HiPIMS that applies variable bias voltage. \citep{nedfors18:031510} We consider a fully ionized deposition flux as a characteristic of HiPIMS deposition in the MD framework. Thus, the film density, surface roughness and microstructure as well as film-substrate interface are probed during film deposition with atomic resolution.
%----------------------------------------------------------------------------
\section{Method}

The detailed description of the simulation procedure can be found in our earlier work. \citep{kateb2019} Briefly, MD simulations \citep{allen1989} were performed using the LAMMPS \footnote{LAMMPS website, \url{http://lammps.sandia.gov/}, distribution 14-April-2018} open source code. \citep{plimpton1995,plimpton2012}

We assume thermal evaporation to have completely neutral flux and use a fully ionized flux to represent HiPIMS. We neglect the working gas in our simulation since its effect on the film properties is minor. \citep{fang1993} Besides, the ratio of working gas ions in the HiPIMS process can be insignificant, e.g. in the case of Cu it has been reported that up to 92\% of ions arriving at the substrate are ions of the target material. \citep{vlcek07:45002}

The importance of the various potential energy terms for energetic flux of ions are described by \citet{anders02:1100}. The combination we used here almost entirely follows his suggestion except for those terms defined by the contribution of electrons. In particular, the image charge acceleration term is only important for dielectric surfaces and can be neglected for the case of copper. Also it is not possible to consider electronic excitation and relaxation in our scheme as the electron force field has only been developed up to the 3rd row of the periodic table and is not available for the transition metals. \citep{su2007,jaramillo2011} The interactions of film/substrate atoms and neutral flux atoms was modeled using the embedded-atom method (EAM) force field. \citep{daw1983,daw1984} The total potential energy of atom $i$, $E_i$ is described by
\begin{equation}
	E_i=F_i(\rho_i)+\frac{1}{2}\sum_{i\neq j}\phi_{ij}(r_{ij})
\end{equation}
where $F_i$ is the embedding energy of atom $i$ into electron density $\rho_i$ and $\phi_{ij}$ is a pair potential interaction of atom $i$ and $j$ at distance $r_{ij}$. 

The multi-body nature of the EAM potential is a result of the embedding energy term i.e.\ $\rho_i$ itself depends on electron density of neighboring atoms $\rho_{ij}$
\begin{equation}
	\rho_{i}=\sum_{i\neq j}\rho_{ij}(r_{ij})
\end{equation}

The ion-ion interaction in the flux was modeled via a hybrid approach based on EAM and Ziegler-Biersack-Littmark (ZBL) \citep[Chap.~2]{ziegler1985} potential. The latter takes into account both short range Coulombic interaction and long range screening
\begin{equation}
	V(r_{ij})=\frac{Z_iZ_je^2}{4\pi\varepsilon_0r_{ij}}\Phi\left(\frac{r_{ij}}{a}\right)
\end{equation}
where the $Z_i$ and $Z_j$ are the atomic numbers of species $i$ and $j$ that belong to Coulombic term and $e$ and $\varepsilon_0$ stand for elementary charge and vacuum permitivity, respectively.

The universal screening function in reduced unit is defined 
\begin{equation}
	\Phi\left(\frac{r_{ij}}{a}\right)=\sum_{n=1}^4a_n e^{-c_nr_{ij}/a}
\end{equation}
where $a$ is the ZBL modification of Bohr's universal reduced coordinate with 0.8853 derived from Thomas-Fermi atom
\begin{equation}
	a=\frac{0.8853a_0}{Z_i^{0.23}+Z_j^{0.23}}
\end{equation}
with $a_0$ being Bohr radius and $a_n$ is normalizing factor i.e. $\sum a_n=1$.
\begin{eqnarray}
    &&a_n= 0.18175, 0.50986, 0.28022, 0.02817\nonumber\\
    &&c_n= 3.19980, 0.94229, 0.40290, 0.20162\nonumber
\end{eqnarray}

The ionization energy is described by the ZBL potential. This, may lead to exaggerated etching due to the repulsive force of the ZBL potential. We solve this issue by using a checking mechanism for neutralized ions, i.e. if an ion stays on the surface or implants into sublayers and remains there for 1~ps, it is considered an atom belonging to the film and its ZBL potential is turned off. This 1~ps is chosen to be consistent with the time scale for electron and lattice to reach an equilibrium state after localized heating e.g. due to high energy ion impacts. \cite{anders02:1100}

The substrate was considered to be a single crystal Cu with 77$\times$90~{\AA}$^2$ lateral dimensions and its (111) plane being exposed to the deposition flux which makes the growth direction to be along the $\langle111\rangle$ orientation. The initial configuration consisted of a rigid monolayer, three monolayers as a thermostat layer and 12 monolayers as a surface layer. The initial velocities of substrate atoms were defined randomly from a Gaussian distribution to mimic temperature of 300~K and the substrate energy was minimized and relaxed afterwards.

For all cases, the deposition flux consisting of 22000 atoms was introduced at a distance of 150~nm above the substrate surface. The initial velocity of ions towards substrate were assigned randomly with a flat distribution within 0 -- 40, 60 --100 and 120 -- 160~eV energy ranges. Keep in mind that a minimum displacement energy, to dislodge an atom in the substrate, is in the range of 10 -- 40~eV, so that the two higher energy regimes imply that the bombarding species energy exceeds the binding energy and the surface binding energy of the growing film. The process of introducing species was a single atom/ion every 1~ps which produces an equal deposition rate in all cases. This deposition rate is one order of magnitude smaller than in our previous study. \citep{kateb2019} We would like to remark that the length of voltage pulses in a HiPIMS deposition is normally 50 -- 400 $\mu$s \citep{gudmundsson12:030801} which is much longer than the simulation times achieved by MD simulation. \cite{kateb2012} Here the impulse nature of HiPIMS was neglected and main attention has been brought to the effect of ionization. Note that the higher energy regimes, when fully ionized, can also be thought of as mimicking a cathodic arc deposition. \citep{anders02:1100} As has been shown earlier a highly ionized flux is able to capture several aspects of HiPIMS deposition. \citep{kateb2019}

The time integration of the equation of motion was performed regarding the microcanonical ensemble (NVE) with a timestep of 5~fs. The Langevin thermostat \citep{schneider1978} was only applied to the thermostat layer with a damping of 5~ps for a total time of 25~ns. The thermostat layer is responsible for heat dissipation \citep{srivastava1989} and damping defines timescale for this purpose.

The most common structure characterization is to utilize the radial distribution function, $g(r)$. \citep{kateb2018b,azadeh2019} In order to study the film-substrate interface quantitatively, we utilized partial $g(r)$, $g_{ij}(r)$. This approach was originally introduced by \citet{ashcroft1967} for analyzing binary mixtures. However, $i$ and $j$ in our study are defined to distinguish between film and substrate atoms. Knowing the atomic coordinates of a system, $g_{ij}(r)$ can be defined as
\begin{equation}
    g_{ij}(r)=\frac{\rho_{ij}(r)}{\rho_0c_{j}}
	\label{eq:gij}
\end{equation}
where $\rho_{ij}(r)$ is the number of $j$ species in a spherical shell of radius $r$ from the central particle of type $i$, $\rho_0$ is the average number (not mass) density of both species and $c_j$ is the molar fraction of species $j$. The reduced value $g_{ij}(r)$ then can be determined by
\begin{equation}
    G_{ij}(r)=4\pi\rho_0r[g_{ij}(r)-1]
\end{equation}

It is worth mentioning that Eq.~(\ref{eq:gij}) is mainly developed to study the diffraction pattern of homogeneous binary alloys. Thus, by definition $g_{ij}(r)=g_{ji}(r)$ since $\rho_{ij}(r)$ is linearly proportional to $c_j$. Eq.~(\ref{eq:gij}) is defined for a homogeneous mixture and one has to determine a homogeneously mixed region at the interface and then calculate $c_j$ and $g_{ij}(r)$. The homogeneity issue becomes more pronounced when the thickness of the mixed region becomes a few monolayers. Also, the results can be affected depending on how one determines the mixed region. However, assuming a constant $c_j$ would allow the intermixing to be reflected through $\rho_{ij}(r)$ but it leads to $g_{ij}(r)\neq g_{ji}(r)$. We assumed $c_j=0.5$, the ratio of deposition flux to total atoms, that allows comparison of intermixing at different levels.

Although one can determine average structure of a system from $g(r)$, it gives limited information on the local structure. \citep{steinhardt1983,kelchner1998,tsuzuki2007,kateb2018b} To this end, we utilize common neighbor analysis (CNA) which has been shown to be promising for the local structure characterization. Besides, it allows distinction between fcc and hcp which is of practical importance in defect analysis. \citep{kateb2018b,azadeh2019} Briefly, the CNA determines local crystal structure based on decomposition of 1st nearest neighbors (NNs), from 1st $g(r)$ peak, into different angles. Thus, CNA is sensitive to angles between pairs of NNs and can distinguish between fcc and hcp structure. Consequently a twin grain boundary can be determined based on slight angle difference between 1st NNs while it has the same number of NNs at same distance as an fcc atom. The Ovito package\footnote{Ovito website, \url{http://ovito.org/}, Version 2.9.0} was used to generate atomistic illustrations including CNA and $G_{ij}(r)$. \citep{stukowski2009}

%-----------------------------------------------------------------------------
\section{Results and discussion}
\subsection{Interface mixing}

\begin{figure}
    \centering
	\subfigure{\includegraphics[width=1\linewidth]{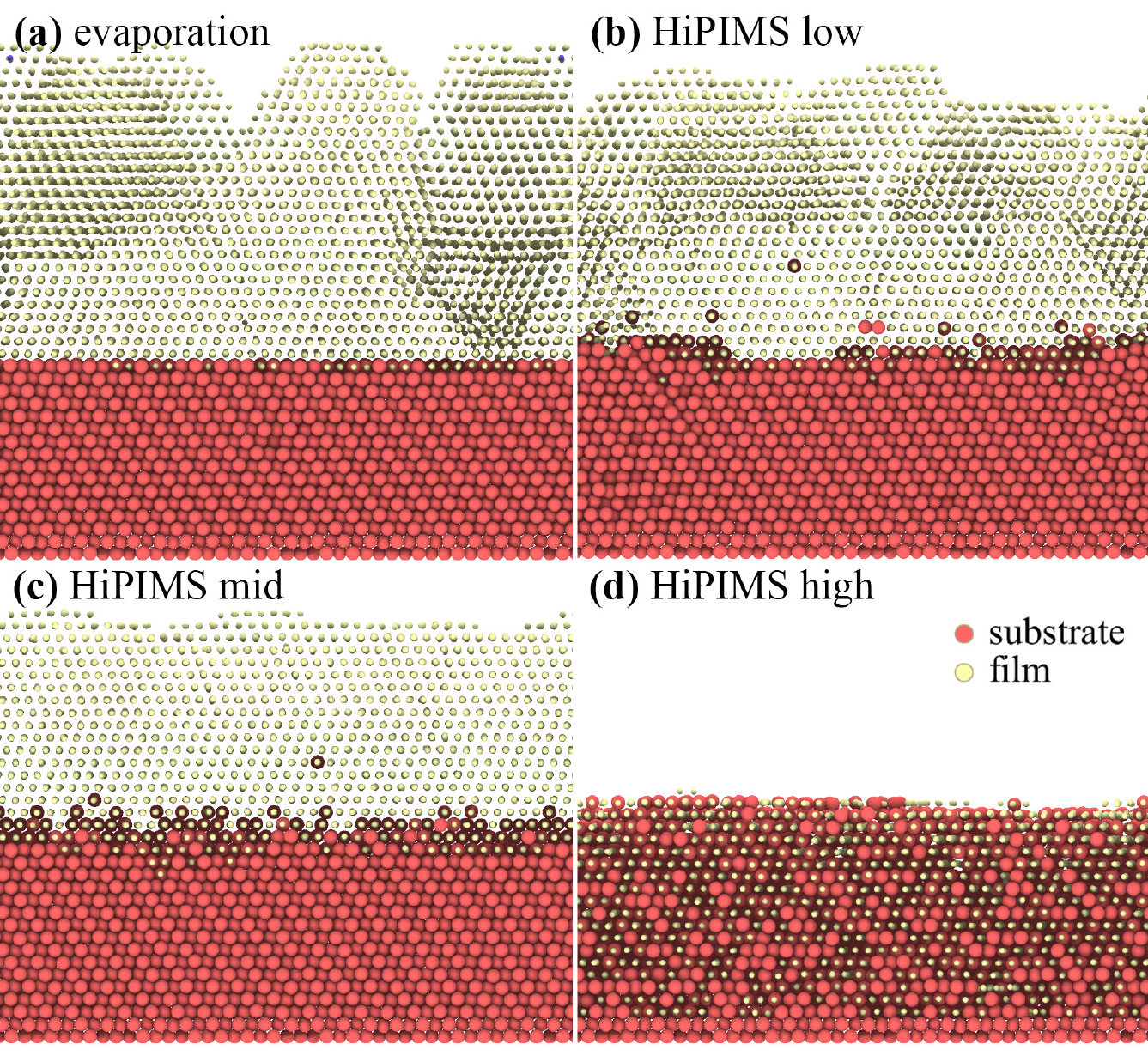}}
    \caption{Illustration of interface mixing after 25~ns deposition using (a) thermal evaporation and HiPIMS with (b) low, (c) moderate and (d) high energy. To distinguish film/substrate intermixing, the film atoms are illustrated with smaller diameter. Note that substrate atoms inside the film appear as black circle.}
    \label{fig:mix25ns}
\end{figure}

Fig.~\ref{fig:mix25ns} shows the cross section of Cu films deposited under different conditions on identical substrates. For thermal evaporation and low energy HiPIMS, shown in Fig.~\ref{fig:mix25ns}(a) and (b), respectively, different interface mixing is observed. We would like to remark that a similar energy range is utilized for these films and their difference is limited to fully neutral and fully ionized flux in evaporation and HiPIMS, respectively. Thus, ionization of the depositing species in HiPIMS is responsible for the interface mixing. It is worth mentioning that, interface mixing is not a direct act of ions e.g.\ implantation of high energy ions through the film and into the substrate. On the contrary, it has been shown to be sensitive to the temporal lattice excitations localized in the vicinity of impacts using a neutral flux. \citep{sprague1996} More recently, it has been found to be a consequence of bi-collision events \cite{kateb2019} i.e.\ a high energy ion impact that causes amorphization followed by another impact that leads to recrystallization. During the bi-collision event atoms at the interface are stirred and that leaves a mixed interface. Such events become dominant during deposition with higher ratio of ionized flux. \cite{kateb2019} A comparison between Fig.~\ref{fig:mix25ns}(a) and (b) clearly shows the importance of ionized flux in the increase of interface mixing even using a low energy flux. This is of importance when making high quality electrical contacts where limited interface mixing is required but higher energy flux cannot be achieved using evaporation, or in the case of dcMS higher energy is achieved by higher power that also increases deposition rate. Surprisingly, the moderate energy HiPIMS, Fig.~\ref{fig:mix25ns}(c), shows a similar interface mixing to the low energy counterpart. We explain this further in the discussion in $\S$~\ref{micro}. Finally, Fig.~\ref{fig:mix25ns}(d) shows an extreme intermixing between film and substrate in the case of energetic bombardment of the substrate by the film forming species. As demonstrated by MC simulation even an ion with a few hundreds of eV energy can only penetrate to a depth of a few nm unless the film density is less than 80\% of theoretical density. \citep{muller1986,muller1986jap} In an energetic process such as cathodic arc deposition, ions penetrate a few monolayers at the cost of 100~eV/nm energy loss. \citep{anders02:1100} In Fig.~\ref{fig:mix25ns}(d), we observe a large number of film atoms deep into the substrate while we do not have that many ions with hundreds of eV energy. Thus, the presence of film atoms deep into the substrate is associated with densities lower than 80\% that is achieved by etching of the substrate and resputtering of the film. Besides, a severe interface mixing requires both ionized flux and high energy to be realized as it does not occur in low and moderate energy HiPIMS.

\begin{figure}
    \centering
	\subfigure{\includegraphics[width=1\linewidth]{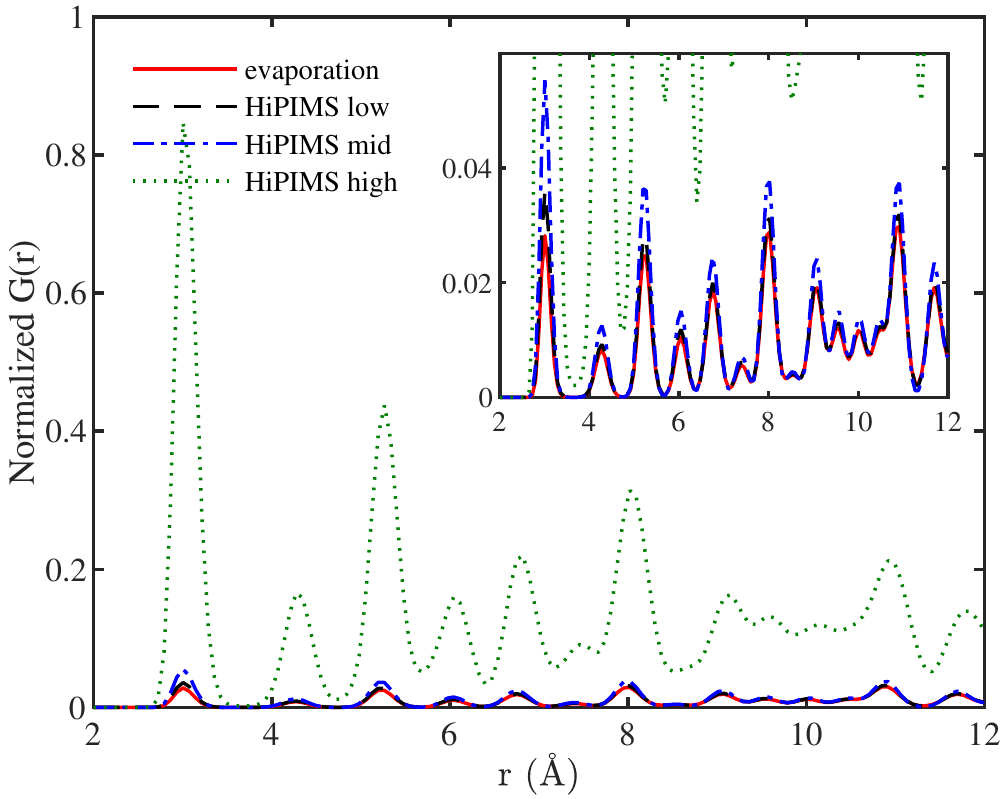}}
    \caption{Variation of film-substrate $G(r)$ as a representer of interface mixing at different condition.}
    \label{fig:gr}
\end{figure}

Furthermore, we explored the distances between substrate and film atoms using $G(r)$ which can represent statistical quantity of interface mixing. Fig.~\ref{fig:gr} shows the $G_{ij}(r)$ patterns obtained after deposition at different conditions using a 12~{\AA} cutoff and 200 bins. Regardless of intensities all patterns show identical number of peaks at similar positions. The first peak at 3~{\AA} is associated with the 1st NNs distance or minimum distance between film-substrate atoms. This peak is highest for high energy HiPIMS which means that the number of film-substrate atom pairs is much higher ($\sim$16 times) than in the rest of the films. As expected, HiPIMS with moderate and low energy present slightly higher intensity than evaporation.  

It is worth mentioning that $G(r)$ can be translated to the structure factor ($S(q)$) \citep{kateb2018b} which can be determined experimentally using X-ray and neutron scattering techniques. However, in order to compare the results the total reduced radial distribution function has to be determined. \citep{sietsma1991}

\subsection{Surface roughness}
\begin{figure}
    \centering
	\subfigure{\includegraphics[width=1\linewidth]{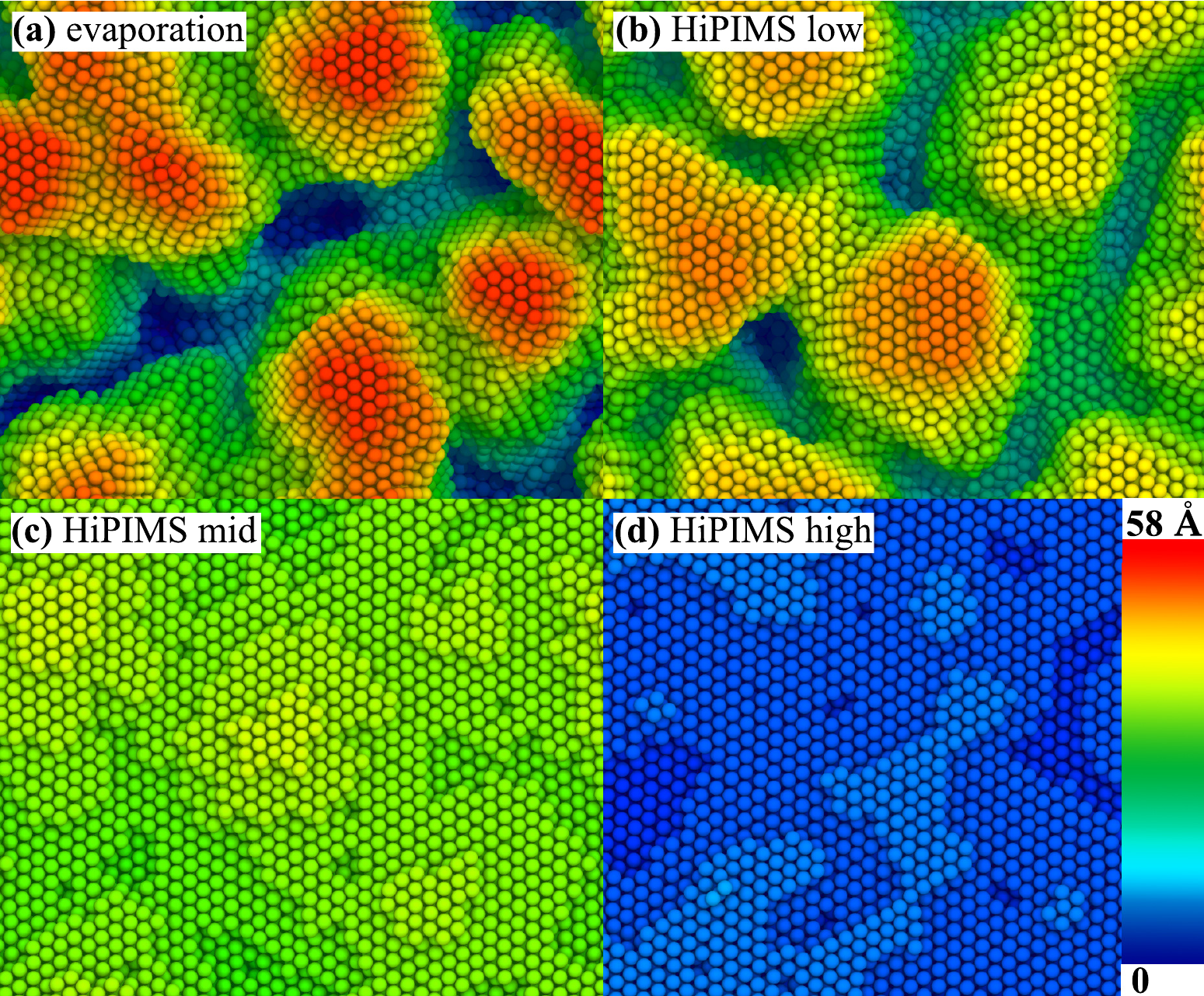}}
    \caption{The surface topology obtained after 25~ns deposition using (a) thermal evaporation and HiPIMS with (b) low, (c) moderate and (d) high energy. The colorbar indicates hight from the substrate surface.}
    \label{fig:Z}
\end{figure}

Fig.~\ref{fig:Z} shows a top view of the films deposited under condition that mimic thermal evaporation and HiPIMS with identical deposition rate and time. The HiPIMS condition are then assumed to have low, medium and high substrate bias. As indicated by the color bar, the dark blue here shows the substrate surface and atoms that are 58~{\AA} above the substrate are identified by red. It can be clearly seen that the films deposited by evaporation and low energy HiPIMS presents considerably rougher surfaces compared to the other cases. However, low energy HiPIMS present slightly smoother film and more coverage than evaporation. This is due to the fact that during thermal evaporation neutral atoms are allowed to form clusters before arriving at the surface while in low energy HiPIMS, the repulsion between ions does not allow clustering and the islands are the result of the energy distribution i.e.\,some adatoms have higher energy and diffuse longer than the others which is responsible for island growth. \citep{kateb2019} Gas phase clustering has already been demonstrated using multi-scale MD simulation i.e. when a target to substrate distance similar to that of a typical experiment is defined and gas phase collisions are treated properly. \citep{brault2018} It is worth mentioning that such an island growth is not an artificial effect of the short time scale and consequently high deposition rate of MD simulations. As a matter of fact, island growth has been reported for deposition of Cu on Cu with experimental rate and modeling the diffusion process through accelerated dynamic simulation. \citep{hubartt2013} In moderate and high energy HiPIMS, bi-collision event \citep{kateb2019} preserves a smooth surface and prohibits island growth. It causes localized amorphization that fills the gaps between islands with an atomically flat surface. Normally secondary collision of energetic ions cause recrystallization of amorphous regions while the smooth surface topology is maintained.

Looking at the similarities between evaporation and low energy HiPIMS, one may think bi-collision events do not occur for low energy HiPIMS and become dominant above a specific energy threshold. We would like to remark that island growth appeared to be direct consequence of having flat energy distribution \citep{kateb2019} and have not been observed previously using monodispersed energy flux. \citep{gilmore1991,dong1998} Thus, in an ionized PVD interplay between surface diffusion and bi-collision events determines the surface roughness. \citep{kateb2019} However, as energy distribution shifts to higher values the number of high energy ions is increased that leads to higher probability of bi-collision events. Thus, for low energy HiPIMS surface diffusion is dominating over bi-collision events and that results in island growth. At moderate energy HiPIMS, the bi-collision event becomes dominant and leaves a smooth surface. Higher energies, however, make etching and resputtering more pronounced while bi-collision events still maintain the surface smoothness.

We also utilized the construct surface mesh (CSM) algorithm \citep{stukowski2014} implemented in OVITO to determine the surface area ($A$). Table~\ref{tab:surfarea} summarizes the CSM result including volume change before and after deposition ($\Delta V$). It is worth mentioning that CSM determines the volume of the solid without its porosities and thus cannot be used to determine the film density. It can be seen that $A$ slightly decreases from evaporation to low energy HiPIMS while its reduction is considerable for moderate energy HiPIMS. For high energy HiPIMS the $A$ is slightly lower than at moderate energy. However, at this stage $A$ is irrelevant since $\Delta V$ obtained using high energy HiPIMS is less than one fifth of the other films.

\begin{table}[h]
\caption{\label{tab:surfarea} Values of film surface area and volume change obtained by CSM. $N_{\rm Film}$ and $N_{\rm Sub}$ denote final number of film and substrate atoms, respectively.}
\begin{tabular}{ c c c c c }
\hline \hline
Method & evaporation & \multicolumn{3}{c}{HiPIMS}\\
Energy & low & low & mid & high \\
\hline
$A$ ({\AA}$^2$)& 18209 & 14515 & 7327 & 7244 \\
$\Delta V$ ({\AA}$^3$)& 254481 & 258620 & 261053 & 49647 \\
$N_{\rm Film}$ (atom)& 22000 & 21993 & 21998 & 15033 \\
$N_{\rm Sub}$ (atom)& 19276 & 19268 & 19275 & 7897 \\

\hline
\end{tabular}
\end{table}

\subsection{Film density}
\begin{figure}
	\centering
    \includegraphics[width=1\linewidth]{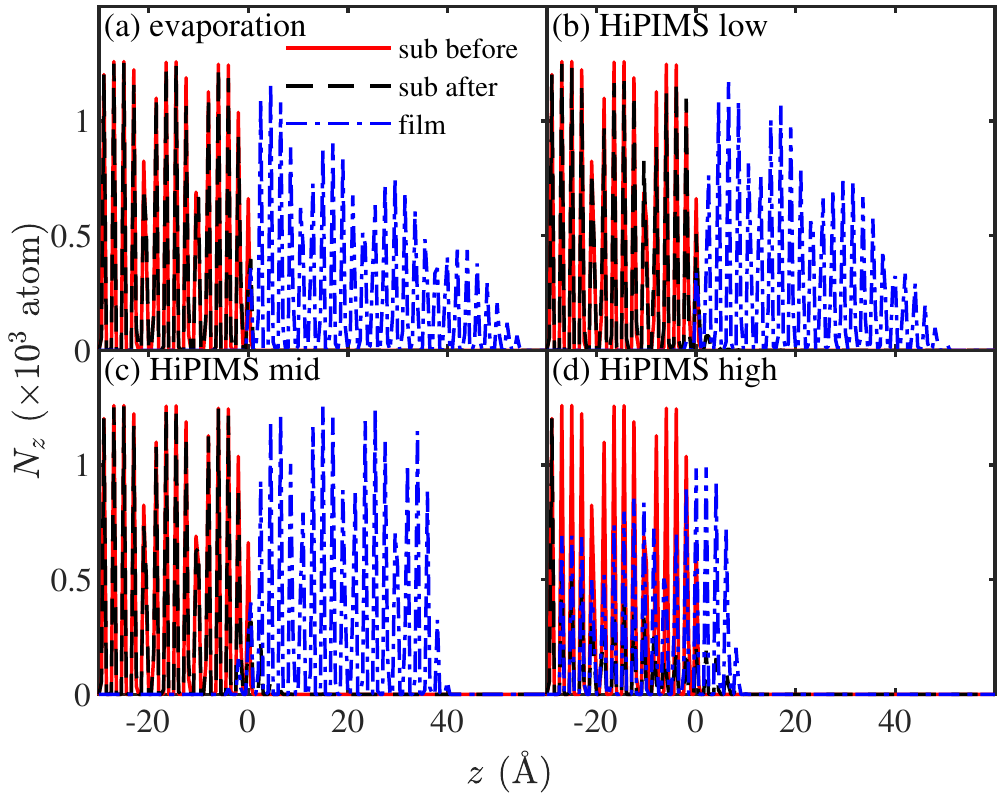}
    \caption{Spatial distribution of atoms along the deposition direction ($N_z$) before and after deposition using (a) thermal evaporation and HiPIMS with (b) low, (c) moderate and (d) high energy. The $z\leq0$ indicates location of substrate atoms before deposition.}
    \label{fig:rho}
\end{figure}

Fig.~\ref{fig:rho} shows the spatial distribution of atoms along the growth direction ($N_z$) which represents film density in atomistic simulations. Before deposition, substrate atoms are indicated by red line (sub before) are located at $z\leq0$. The film and substrate patterns after deposition are also shown by blue and black (film and sub after), respectively. Before deposition, there is a sharp transition at $z=0$ that is an indication of a flat substrate surface. A similar pattern is obtained for the film deposited at moderate energy HiPIMS as seen in Fig.~\ref{fig:rho}(c) which is a signature of layer-by-layer growth. \citep{muller1987prb} On the other hand, both evaporation and HiPIMS deposition using low energies, shown in Fig.~\ref{fig:rho}(a) and (b), decay gradually with the $z$ due to surface roughness as a characteristic of island growth. \citep{gilmore1991} Comparing the substrate before and after deposition, it can be seen that the substrate pattern is mostly preserved except in the case of high energy HiPIMS deposition, shown in Fig.~\ref{fig:rho}(d). In this case, number of substrate atoms decreased after deposition and a significant number of film atoms can be found in $z\leq0$. This supports our claim that for a severe intermixing low surface density is required which is achieved by etching and resputtering of the substrate and film, respectively.

\subsection{Microstructure}
\label{micro}
\begin{figure}
    \centering
	\subfigure{\includegraphics[width=1\linewidth]{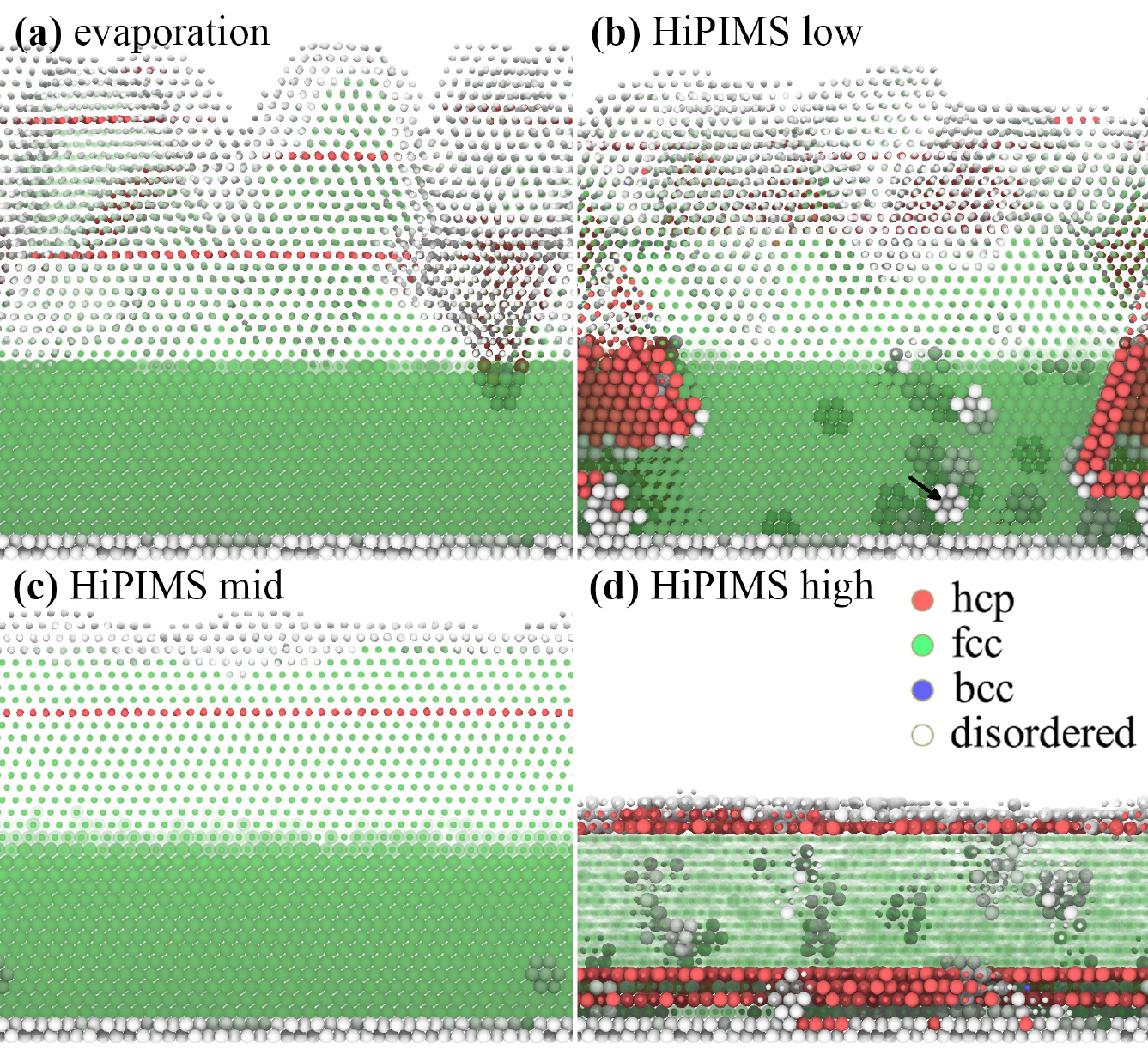}}
    \caption{Variation of local structure obtained after 25~ns deposition using (a) thermal evaporation and HiPIMS with (b) low, (c) moderate and (d) high energy. To distinguish between film/substrate, the film atoms are illustrated with smaller diameter. The fcc atoms are shown transparent to illustrate defects properly. The black arrow in (b) indicates a vacancy surrounded by disordered atoms.}
    \label{fig:micro}
\end{figure}

The microstructures obtained after deposition under different conditions are shown in Fig.~\ref{fig:micro}(a) -- (d). The color contrast obtained by adoptive CNA which can distinguish between different crystal structures i.e.\ fcc, hcp, bcc and disordered atoms indicated by green, red, blue and white, respectively. It can be seen that all films consist of single crystal Cu aside from stacking faults (twin boundaries) and point defects.  Epitaxial growth of single crystals using HiPIMS has been already demonstrated experimentally for fcc elements \citep{cemin2017} and alloys \citep{kateb2018} even with a large lattice mismatch. The formation of stable twin boundaries is a common issue for the Cu deposited by evaporation and HiPIMS. \citep{kateb2019} It has been observed even using accelerated MD simulation with a more realistic diffusion. \citep{hubartt2013} It has also been verified experimentally by polar mapping of the (111) planes in the epitaxial Cu deposited by thermal evaporation \citep{chen2013} and HiPIMS. \citep{cemin2017} It can be seen that low energy HiPIMS deposition presents similar film to that of evaporation. Strictly speaking however, much more defects are introduced into the substrate using low energy HiPIMS. Surprisingly, moderate energy HiPIMS presents minimum crystal defects while again at high energies the defects can be found even in the substrate. We have previously attributed defects in the substrate to high energy collisions. \citep{kateb2019} In the case of evaporation, however, defects are the result of cluster impact. Variation of defect concentration with the HiPIMS energy supports our claims about the frequency of high energy collisions. With the proper frequency of collisions the crystal has enough time to recrystallize and heal while higher repetition of collisions causes amorphization before complete recrystallization and the density of defects increases with time. At low energy, however, the amorphization occurs while the secondary impact, that causes recrystallization, occurs too late when a few atomic layers have been added to the film and it is not effective to reconstruct deep down in the substrate.

An early model that described the microstructure of a growing film was introduced by \citet{klokholm1969}. He suggested that in evaporated metallic films, there exist a disordered liquid-like material buried under the advancing film surface and the amount of disordered film decreases with rising temperature. Based on our observations (here and in our previous study \citep{kateb2019}) we did not detect any disordered material inside the film deposited at 300~K.

When discussing energetic deposition, \citet{musil2000} introduced atomic scale heating and claimed it to be responsible for achieving a dense film via increasing surface mobility at generally low bulk temperature. \citet{anders02:1100} indicated that ions in energetic deposition deliver both kinetic and potential energy which both contribute to atomic scale heating. Thus, it becomes the dominating mechanism in processes such as cathodic arc deposition. We have already demonstrated thermal spikes due to high energy collisions. \citep{kateb2019} However, with the bi-collision event it is amorphization and recrystalization that increases density and reduces surface roughness and not increased surface mobility. We would like to remark that \citet{musil2000} was focused on hard and super hard coatings, that present much stronger bonds than the metallic system we studied. Thus, it is reasonable that the result of bi-collision becomes more localized (more short ranged) and shifts towards film surface as bond strength increases. This means atomistic scale heating and bi-collision events can lead to the same result for the hard films.

We would like to remark that densification \citep{muller1987prb,muller1987} and reorientation of grains \citep{dong1998} using Ar$^+$ ion bombardment have been reported previously. However, it has been thought that heavier ions, such as Cu$^+$, exchange a very large momentum and cause destruction or at least amorphization of the substrate. Fig.~\ref{fig:micro}(c) presents a clear evidence showing that one can grow high quality single crystal using heavy ions by correct choice of substrate bias. This is in agreement with the \emph{extended} structure zone model introduced by \citet{anders2010}, that considers bombardment energy. However, he suggests etching occurs when ion energies exceed the cohesive energy of surface atoms by 2--3 order of magnitude. This condition is not achieved in our simulation scheme for Cu with cohesive energy of 3.54~eV. \citep{mishin2001} However, in high energy HiPIMS  59\% of the substrate and 32\% of the film, etched away at 120--160~eV. We refer to this as \emph{partial} etching which can be observed at much lower energy than \emph{complete} etching condition proposed by extended structure zone model.

\section{Conclusion}

Using MD simulations, HiPIMS deposition (with a fully ionized flux) and thermal evaporation (completely neutral flux) is studied. In all cases, a flat energy distribution is utilized. It is shown that the surface roughness is the product of clustering in the vapor phase and island growth on the substrate surface. The former can be reduced by utilizing ionized flux as ions of the same polarity repel each other. However, reducing island growth is more complex and it occurs through so-called ``bi-collision'' of high energy ions. First a high energy ion implants into subsurface layers and causes localized amorphization which fills the gaps between islands. Secondary ion bombardment causes recrystallization and maintains a smooth surface. There are no high energy ions in the thermal evaporation which presents an extremely rough surface. However, during HiPIMS deposition the number of bi-collision events can be controlled by applying bias voltage to the substrate. As a result low energy HiPIMS (representing low voltage bias) presents slightly lower surface roughness than thermal evaporation. However collisions in low energy HiPIMS are more energetic than cluster impacts in evaporation and thus it introduces more defects into the substrate. Shifting the energy distribution of ions bombarding the substrate, increases the number of high energy ions and the probability of bi-collision events. Thus the bias voltage must be carefully tuned to achieve high quality single crystalline growth. For instance, minimum surface roughness and crystal defects are achieved at moderate energy HiPIMS that represents moderate bias voltages. High energy HiPIMS, however, enters into etching regime due to high frequency of bi-collision events.  

\section*{Acknowledgments}
This work was partially supported by the University of Iceland Research Funds for Doctoral students, the Icelandic Research Fund Grant Nos.~196141, 130029 and 120002023.

\bibliographystyle{aipnum4-1}
\bibliography{ref}
\end{document}